\documentclass[twocolumn,preprintnumbers,amsmath,amssymb,prb,floatfix,superscriptaddress,nofootinbib,longbibliography]{revtex4-2} 
\usepackage{epic}
\usepackage{graphicx}
\usepackage[caption=false]{subfig}
\usepackage[T1]{fontenc}
\usepackage{stackrel}
\usepackage{float}
\usepackage{lipsum}   
\usepackage{amsmath,amssymb}
\usepackage{enumitem}  
\usepackage{array}
\usepackage{amssymb,amsmath,amsfonts}
\usepackage{braket}
\usepackage{soul}
\newcommand{\be}{\begin{equation}}
\newcommand{\ee}{\end{equation}}
\newcommand{\bse}{\begin{subequations}}
\newcommand{\ese}{\end{subequations}}
\newcommand{\ba}{\begin{eqnarray}}
\newcommand{\ea}{\end{eqnarray}}

\usepackage{color}
\usepackage[normalem]{ulem}  
\usepackage{float}

\usepackage{hyperref}
\hypersetup{colorlinks=true, citecolor=blue, urlcolor=blue, linkcolor=blue}

\usepackage[margin=1in]{geometry}

\newcommand{\psiu}{\psi^{\uparrow}}

\newcommand{\na}{\nabla}
\newcommand{\psid}{\psi^{\downarrow}}

\newcommand{\Vuu}{V^{\uparrow \uparrow}}
\newcommand{\Vdd}{V^{\downarrow \downarrow}}
\newcommand{\Vdu}{V^{\downarrow \uparrow}}
\newcommand{\Vud}{V^{\uparrow \downarrow}}

\newcommand{\uu}{\uparrow \uparrow}
\newcommand{\dd}{\downarrow \downarrow}
\newcommand{\ud}{\uparrow \downarrow}
\newcommand{\du}{\downarrow \uparrow}

\usepackage{color}
\usepackage[normalem]{ulem}  

\newcommand{\si}{\sigma}
\newcommand{\la}{\lambda}

\usepackage{lineno}

\begin{document}

\title{
 Quantum Dynamics of Electron Scattering from Skyrmions
} 
\author{Hareram Swain}\altaffiliation[dhareram1993@physics.iitm.ac.in]{}\affiliation{Condensed Matter Theory and Computational Lab, Department of Physics, IIT Madras, Chennai-600036, India}
\affiliation{Center for Atomistic Modelling and Materials Design, IIT Madras, Chennai-600036, India}
\author{Arijit Mandal}{}\affiliation{Condensed Matter Theory and Computational Lab, Department of Physics, IIT Madras, Chennai-600036, India}
\affiliation{Center for Atomistic Modelling and Materials Design, IIT Madras, Chennai-600036, India}
\author{S. Satpathy}\altaffiliation[satpathys@missouri.edu]{}\affiliation{Condensed Matter Theory and Computational Lab, Department of Physics, IIT Madras, Chennai-600036, India}
\affiliation{Center for Atomistic Modelling and Materials Design, IIT Madras, Chennai-600036, India}
\affiliation{Department of Physics \& Astronomy, University of Missouri, Columbia, MO 65211, USA}    
\author{B. R. K. Nanda}\altaffiliation[nandab@iitm.ac.in]{}\affiliation{Condensed Matter Theory and Computational Lab, Department of Physics, IIT Madras, Chennai-600036, India}
\affiliation{Center for Atomistic Modelling and Materials Design, IIT Madras, Chennai-600036, India}

\begin{abstract}

Scattering of electrons from chiral spin textures such as the skyrmions is an emerging research area due to its richness in topological quantum transport, which is significant for spintronic devices. We study the dynamical process of scattering of the spin-$\frac{1}{2}$ particles in the form of Gaussian wavepackets from skyrmions with the aid of the non-relativistic time-dependent Schr\"odinger equation. The scattering cross section shows a rich angular dependence and is deterministically influenced by the iterative flipping of the spin state inside the skyrmion. The latter leads to a set of non-trivial outcomes which include finite transmission and reflection probabilities irrespective of interaction strength, formation of secondary wavefronts associated with back-converted spin components, and a long-lived quasi-bound state at the scattering center. In addition to the rich and intriguing physics, the numerical recipe developed here can be easily adopted for any arbitrary spin texture, which will prepare a playground to explore tunable spin transport.

\end{abstract}

\maketitle

\section{Introduction}\label{Sec:Intro}
Magnetic skyrmions are topologically protected solitons defined through a swirling spin texture with a quantized winding number \cite{skx1, skx2, skx3, skx4}. The metastability, high velocity, and emergent magnetic field of skyrmions prepare a fertile playground for device applications in areas such as spintronics \cite{spintronics}, racetrack memory devices \cite{racetrack}, quantum computation technologies \cite{qubit1, qubit2}, etc. Therefore, in recent years, a great deal of research has been devoted to the transport properties of electrons under the influence of skyrmion, such as topological Hall effect (THE) \cite{the1, the2, the3, the5} and spin Hall effect (SHE) \cite{she1, she2, she3}. They arise due to the transverse motion of electrons, which is generally explained by the emergent magnetic field of the skyrmion \cite{Nagaosa2013, mandal}. However, such an approach is valid only in the adiabatic regime, where the Hund's coupling, $J$, is very high such that the electron spins follow the skyrmion spins during hopping. Since $J$ is a property of the material hosting the skyrmion, the adiabatic approximation is not always applicable, and hence, for the accurate description of electron-skyrmion interaction, an alternative and robust quantum treatment is required to accommodate the full spectrum of $J$.  

Stable skyrmion states have already been observed in a plethora of materials. \cite{tokura-review, skx2, semicon-skx, the-com2, the-com3}. However, natural synthesis processes often lead to sporadic distribution of skyrmions instead of a dense or regular periodic array forming a crystal. Therefore, it is practically more relevant to study the electron propagation in the presence of the spin potential formed by isolated skyrmions, in particular, and non-collinear spin texture (NCS) in general. In this direction, so far, the Lipmann-Schwinger(L-S) theory of quantum scattering is explored  \cite{tatara, Denisov_2016, Denisov_2017, blugel}. 

\begin{figure}[hbt!]
    \centering
    \includegraphics[width=0.8\linewidth]{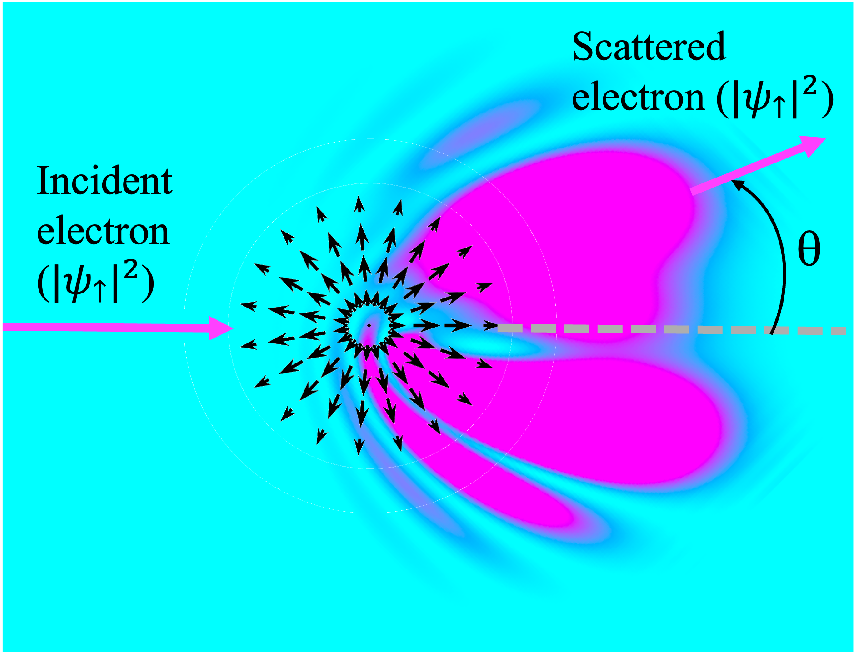}
    \caption{ Electron scattering from an isolated skyrmion obtained from the solution of the time-dependent Schr\"odinger equation. The color coding indicates the scattered wave packet at a fixed time after the collision.}
    \label{fig:problem figure}
\end{figure}

In this static theory, one obtains the final state of the scattered electrons far away from the potential by solving the L-S equation, and thus is unable to capture the information hidden in real-time propagation. Furthermore, the L-S equation is solved either by approximating the scattering to the second Born approximation, which is possible only in the low $J$ regime, or exactly by incorporating the partial wave analysis \cite{Denisov_2017}, which demands continuous symmetry of the potential. Therefore, a large class of NCS that has a higher $J$ and/or lacks continuous symmetry (e.g., bimeron)  can not be investigated using the existing approaches. Therefore, a generalized scattering theory in the presence of a spatially varying potential, including the spin degrees of freedom, is yet to be developed via the real-time propagation of electrons.

A novel and increasingly powerful approach to studying electron scattering in the presence of NCS involves the numerical solution of the time-dependent Schrödinger equation (TDSE) for an evolving quantum wavefunction. Unlike static approaches, which compute only the final state (i.e, far from the scatterer), the TDSE captures the dynamical interference effects, spin conversion, and both adiabatic and non-adiabatic transitions. In addition, this method allows direct visualization of wavepacket dynamics, enabling the study of transient phenomena such as quantum reflection, trapping, and resonant tunnelling that are inaccessible through static methods. As a whole, the TDSE offers deeper insight into the dynamics of NCS-driven scattering processes at play and their potential applications in quantum transport and information processing.

In this work, we develop the numerical method of solving the TDSE for an electron wavepacket scattered from an isolated NCS and employ it to study the case of 1D and 2D skyrmions. Stable noncollinear chiral magnetic states in the form of both 1D, such as spin helix \cite{spin-helix} or spin spiral \cite{spin-spiral}, and 2D, such as skyrmion \cite{tokura-review, skx2, semicon-skx, the-com2, the-com3}, have already been observed in a plethora of materials. Stable 1D non-collinear chiral spin textures are experimentally observed. However, here we have considered a 1D skyrmion as a representative NCS to draw a mathematical and physical parallel to its 2D counterpart.  The purpose of investigating skyrmions is two-fold. Firstly, these are the most widely experimentally synthesized NCS, and secondly, there is no prior information on the intermediate quantum states of the propagating electrons. The developed numerical method validates the existing results reported using the static approaches and, most importantly, provides the following key findings, which have not been reported so far.

Electrons exhibit iterative spin flipping within the skyrmion core, observed in the form of secondary and tertiary wavefronts, and play a crucial role in the scattering outcomes. Counterintuitively, the transmission (scattering along the incident direction) in the spin-flip channels is strongly suppressed, even when the interaction strength is low, which occurs when the Hund's coupling is significantly lower than the kinetic energy of the electron. 

For the 2D skyrmion system, the scattering cross section of the spin-preserving channel saturates at high interaction strength with transmission probability asymptotically approaching zero. Correspondingly, in the same limit, for the 1D skyrmionic system, non-trivially, both transmission and reflection probabilities remain finite. Further, localized metastable spin-down states are dynamically generated via spin-flipping near the skyrmion core. These results highlight the importance and ability of the TDSE approach to scan the transient phenomena that control the final scattered state in the universal class of NCS.

\section{Theoretical Formulation}\label{Sec:Theoretical Formulation}
The objective of this study is to develop a microscopic understanding of the scattering process that arises from the interaction between an electron and a magnetic skyrmion. To achieve this, we present here the theoretical framework to obtain the real-time evolution of the electron wave packet as it propagates through the skyrmion texture, schematically shown in Fig.~\ref{fig:problem figure}.

The spin distribution of a skyrmion at each site is defined through a one-to-one mapping between the location of spin in $\mathbb{R}^2$, represented in polar coordinates with the parameters ($r, \alpha$) and the surface of the Bloch sphere, $S^2$, represented by the parameters ($\theta, \phi$). The mapping of the spins can be represented in linear form as follows:
\begin{align}\label{Eq:skyrmion coordinate}
   &\theta(r) = 
   \left\{
   \begin{array}{ll}
        \left(-\frac{r}{a}+1\right)\pi & \text{for } r \leq a \\
        0 & \text{for } r > a
   \end{array}
   \right.\nonumber\\
   & \phi(\alpha) = m \alpha + \gamma
\end{align}
where $m$ is the winding number, characterizing the topological charge of the skyrmion, and $\gamma \in [0,2\pi)$ is the helicity, controlling the in-plane rotation of the spins. The radius of the disc that accommodates the skyrmion is $a$. For example, in the case of the Néel skyrmion ($m=1$ and $\gamma=0$), Eq. \eqref{Eq:skyrmion coordinate} yields that, the spin points along the $-\hat{z}$ ($\theta =\pi$) at the center of the disc ($r=0$), while it points along $+\hat{z}$ ($\theta=0$) at the boundary ($r=a$). The azimuthal angle $\phi$ wraps around the disc $m$ times as $\alpha$ goes from $0$ to $2\pi$, leading to a topologically nontrivial mapping from the plane to the surface of the Bloch sphere. Outside the skyrmion boundary, the itinerant electron is placed in a homogeneous ferromagnetic background.

\begin{figure*}[t]
    \centering
    \includegraphics[width=\linewidth]{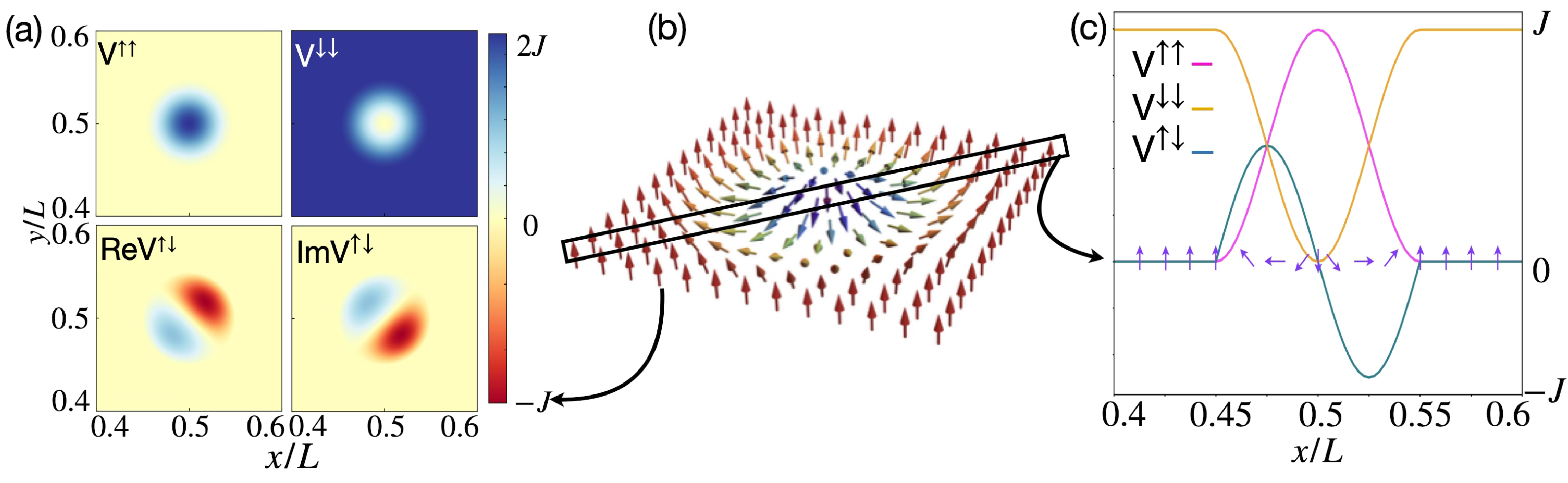}\\
    \caption {(a) The contour map of the potential components for 2D skyrmion based on Eq.~\eqref{Eq:sky pot in}. The color gradient indicates the potential strength. The skyrmion is centered at $(0.5L,0.5L)$ with a radius of $0.05L$ where $L$ is the length of the box. The off-diagonal terms are complex with $\Vud = V^{\downarrow\uparrow*}$. (b) A 3D schematic diagram for a 2D skyrmion. The marked rectangle indicates a 1D cross-section of the 2D skyrmion, which we defined as 1D skyrmion. (c) Potential components of this 1D skyrmion centered at $0.5L$ with length $0.1L$. 
    }
    \label{Fig:sky_pot}
\end{figure*}

The electron interacts with the magnetization field of the NCS via Hund's exchange interaction:
\begin{align}\label{Eq: Hund's exchange interaction}
&   V(\mathbf{r}) = -J \,\, (\mathbf{S} (\mathbf{r}) \cdot \boldsymbol{\sigma})+J\,\,\mathbb{I}
\end{align}
where, $J$ is the exchange interaction strength, $\mathbf{S}(\mathbf{r}) =(\sin{\theta(r)}\cos{\phi(\alpha)}, \sin{\theta(r)}\sin{\phi(\alpha)},\cos{\theta(r)})$ is the local magnetization vector, $\boldsymbol{\sigma} =(\si_x,\si_y,\si_z)$ is the Pauli spin operator, and $\mathbb{I}$ is the identity operator. In Eq.~\eqref{Eq: Hund's exchange interaction} $J\,\, \mathbb{I}$ is an additive constant which ensures that the propagating spin-up electron outside the skyrmion boundary experiences zero potential while the spin-down electron experiences a potential of $2J$. Projecting this electron-skyrmion effective potential in $\sigma_z$ basis we get,
\begin{align}
 V(r,\alpha) & =  
   \begin{pmatrix}
     \Vuu & \Vud\\
     \Vdu & \Vdd
    \end{pmatrix} \nonumber\\
&  = -J\begin{pmatrix}
     \cos{\theta(r)}-1 & \sin\theta(r)\,e^{-i\phi(\alpha)}\\
     \sin\theta(r)\,e^{i\phi(\alpha)}  & -\cos\theta(r)-1
   \end{pmatrix} \label{Eq:sky pot in} 
\end{align}                       
where $\theta(r)$ and $\phi(\alpha)$ describe the local magnetization direction, as introduced in Eq. \eqref{Eq:skyrmion coordinate}. The contour plots of different matrix elements of $V(r,\alpha)$ are shown in Fig.~\ref{Fig:sky_pot}(a).  Fig.~\ref{Fig:sky_pot}(b) represents the potential generated by the 1D skyrmion and is shown as violet arrows in the same figure. The full Hamiltonian for the propagating electron is:
\begin{align}\label{Eq:Full Hamitonian of the electron}
    \mathcal{H} = \mathcal{H}_0 +  V(\mathbf{r}), \quad \quad \mathcal{H}_0 = \frac{\hat{\mathbf{p}}^2}{2m_e}.
\end{align}

The incident electron located at $r_0$ at time $t=0$ with spin pointed along $(\Theta, \Phi)$ is represented by the following normalized Gaussian wave packet of width $\sigma_0$  

\begin{align}\label{Eq:Gaussian wave packet}
\psi(\mathbf{r},0)&\equiv \begin{pmatrix}
                       \psiu (\mathbf{r},0) & \\
                       \psid (\mathbf{r},0)
                     \end{pmatrix}\nonumber\\
                     & =\sigma_0^{-d/2}\pi^{-d/4} e^{\left(i\mathbf{k}_0 \cdot \mathbf{r}-\frac{|\mathbf{r}-\mathbf{r}_0|^2}{2\sigma_0^2}\right)} \begin{pmatrix}
                       \cos(\frac{\Theta}{2}) & \\
                       \sin(\frac{\Theta}{2}) \,e^{-i\Phi}
                     \end{pmatrix}
\end{align}

The time evolution of the wave packet ($\psi(\mathbf{r},t)$) is governed by TDSE, which we solve numerically using the method outlined in section \ref{sec: 2D numerical}, and the resulting scattering dynamics are analyzed in detail in section
\ref{subsec:results}. While the choice of initial wave function is in principle arbitrary, the Gaussian form is adopted for computational convenience, owing to its localized nature and well-defined momentum distribution. 

\section{Numerical Recipe}\label{sec: 2D numerical}

\begin{figure*}[ht]
    \centering
    \includegraphics[width=0.9\linewidth]{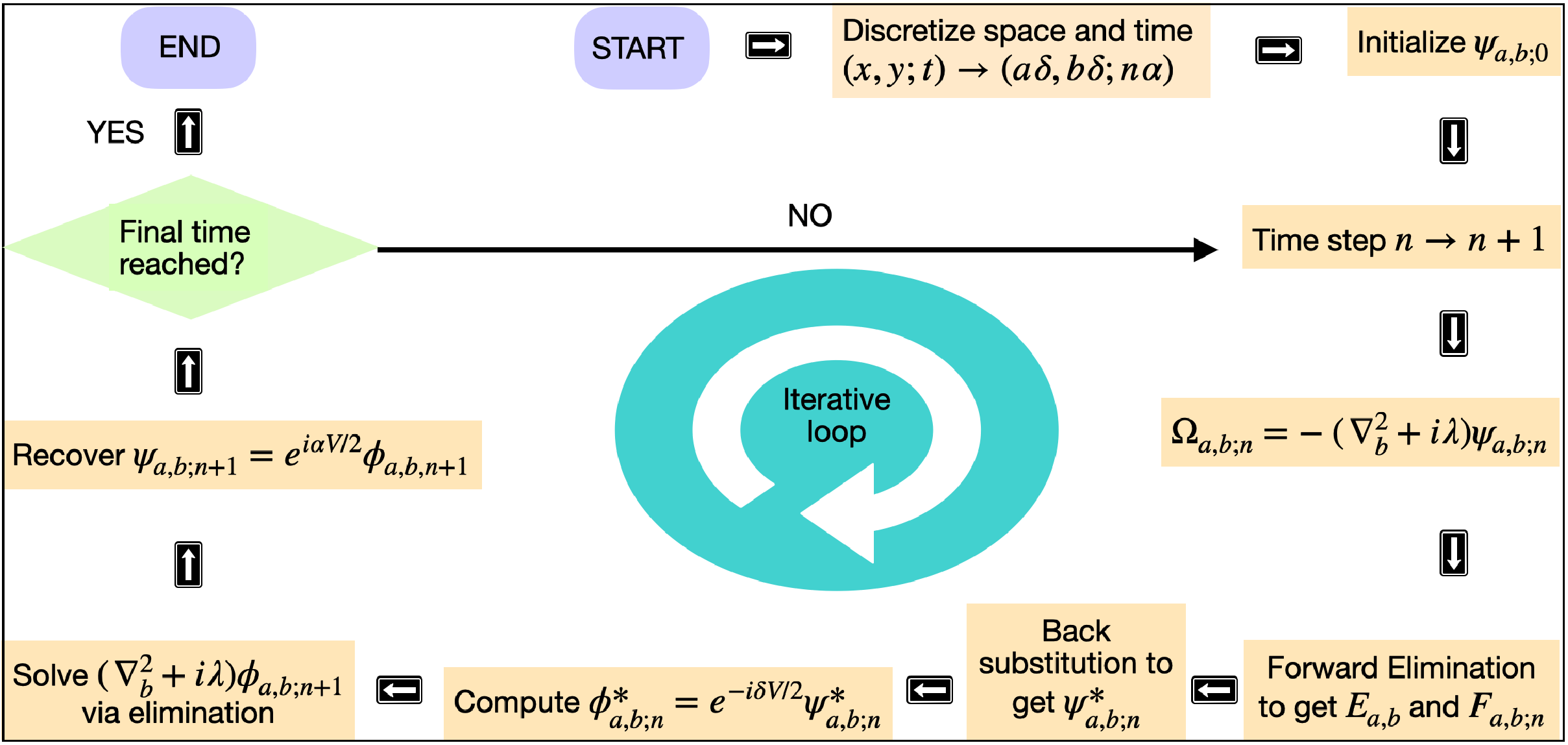}
    \caption{The flow chart of the numerical methods described in Sec.~\ref{Sec:Theoretical Formulation} to solve the TDSE of the propagating electron in any arbitrary 2D potential. A finite difference method is followed, and a hybridized technique of operator-splitting with forward elimination and back substitution is employed. This approach reduces the computational complexity and enhances the numerical stability as explained in Appendix-C, which provides an improvement over the conventional matrix inversion method \cite{10.1119/1.13811}.}
    \label{fig:Flowchart}
\end{figure*}
In this section, we develop the numerical approach to solve the two-dimensional TDSE for the propagating electron wave packet in the presence of a generic spatially varying spin-dependent potential so that the developed method can be used for any arbitrary NCS, including skyrmion. This method is inspired by the work of Galbraith et al \cite{Galbraith1984}; however, our formalism differs from their approach, which relies on direct matrix inversion to get the time-evolved state. As will be elaborated later, here, we adopt an iterative scheme based on forward elimination and back-substitution. This strategy avoids the computational overhead associated with large-scale matrix inversion and offers improved scalability. Specifically, the operator splitting reduces the computational complexity by a factor of $N^4$ ($N \times N$ is space matrix size), making it more efficient for high-resolution grids or real-time propagation over extended durations. The finite-difference scheme, accurate to second order in space and first order in time, is employed here, ensuring stability and accuracy for evolving spinor wave functions $\psi(\mathbf{r},t)$. As it is evident below, the algorithm (see Fig.~\ref{fig:Flowchart} proceeds step-by-step in time, updating the wave function at each spatial grid point based on local potential values and nearest-neighbor couplings.

The TDSE for the spin-$\frac{1}{2}$ particle in two dimension is (in the unit of $\hbar=1$ and mass $m_e=\frac{1}{2}$)
\begin{align}\label{Eq:schrodinger equation for spin 1/2 position basis}
i\frac{d}{dt}\psi(x,y;t) & = \Bigg(\mathcal{H}_0+V(x,y)\Bigg) \psi(x,y;t)\nonumber\\
                         & =\Bigg(-\frac{\partial^2}{\partial x^2}- \frac{\partial^2}{\partial y^2} +  V(x,y) \Bigg) \psi(x,y;t)
\end{align}
Eq.~\eqref{Eq:schrodinger equation for spin 1/2 position basis} is solved numerically by following the finite difference scheme as in \cite{Goldberg1967ComputerGeneratedMP} but for a spin system in two dimensions. To this end, we consider a two-dimensional square mesh discretized in space and time. The spatial coordinates $(x,y)$ are defined on a grid with uniform spacing $\delta$ and time is discretized in steps of size $\alpha$. The wave function in this discretized space and time is expressed as, 
\begin{align}\label{Eq:psi ijn defn}
\psi(x,y;t) \rightarrow\psi_{a,b;n} \equiv \psi(a\delta,b\delta;n\alpha )  
\end{align}
\begin{align}
    0 \leq \lbrace{x,y}\rbrace \leq L; \quad \quad 0\leq \lbrace{a,b}\rbrace \leq \mathbb{J}
\end{align}
where, $L(=\mathbb{J}\delta)$ is the length of the computational box. The time evolution of the state following the TDSE is given as
\begin{align}\label{Eq:formal solution to schrodinger equation dicretized}
    \psi_{a,b;n+1} = e^{-i\alpha \mathcal{H}} \psi_{a,b;n}
\end{align}

As the time evolution operator $e^{-i\alpha \mathcal{H}}$ is unitary, it is important that the same must be preserved while performing the numerical calculations. To this end, $\alpha$ being small, we approximate the exponential using a symmetric rational approximation accurate to $\mathcal{O}(\alpha)$:
\begin{align}\label{Eq: unitary e-i alpha H}
    e^{-i\alpha \mathcal{H}}=e^{-i\alpha (\mathcal{H}_0+V)}\approx \frac{1-\frac{1}{2}i\alpha \mathcal{H}_0+\mathcal{O}(\alpha^2)}{1+\frac{1}{2}i\alpha \mathcal{H}_0+\mathcal{O}(\alpha^2)} \frac{e^{-i\alpha \frac{V}{2}}}{e^{i\alpha \frac{V}{2}}}
\end{align}
Substituting Eq.~\eqref{Eq: unitary e-i alpha H} in Eq.~\eqref{Eq:formal solution to schrodinger equation dicretized} and rearranging yields,

\begin{align}\label{Eq: original finite diff eqn}
   && \left(\na_a^2+i\lambda\right)\left(\na_b^2+i\lambda\right)e^{i\alpha\frac{V}{2}}\psi_{a,b;n+1} = \nonumber\\
    && \left(-\na_a^2+i\lambda\right)\left(-\na_b^2+i\lambda\right)e^{-i\alpha \frac{V}{2}}\psi_{a,b;n}
\end{align} 

where the discrete Laplacians are given by

\begin{subequations}\label{Eq:defn nabla^2_a}
   \begin{align}
     \na_a^2 \psi_{a,b;n} = \psi_{a+1,b;n}-2\psi_{a,b;n}+\psi_{a-1,b;n} \label{Eq:defn nabla^2_a}\\
     \na_b^2 \psi_{a,b;n} = \psi_{a,b+1;n}-2\psi_{a,b;n}+\psi_{a,b-1;n}\label{Eq:defn nabla^2_b}
   \end{align} 
\end{subequations}

and the dimensionless parameter $\lambda$ is defined as
\begin{align}\label{Eq:lambda defn}
    \lambda = \frac{2\delta^2}{\alpha}
\end{align}

Instead of solving Eq.~\eqref{Eq: original finite diff eqn} directly, we employ an operator-splitting technique and introduce an intermediate state $\psi^*_{a,b;n}$, yielding two decoupled equations:
\begin{subequations}\label{Eq:reduced diff eq}
\begin{align}
    &\left(\na_a^2+i\lambda\right)\psi^*_{a,b;n} = \left(-\na_b^2+i\lambda\right)\psi_{a,b;n} \label{SEq:reduced diff eq_a}\\
    &\left(\na_b^2+i\lambda\right)e^{i\alpha \frac{V}{2}}\psi_{a,b;n+1} \nonumber\\
          & \hspace{2.5cm}=\left(-\na_a^2+i\lambda\right)e^{-i\alpha \frac{V}{2}}\psi^*_{a,b;n} \label{SEq:reduced diff eq_b}
\end{align}
\end{subequations}
Eq.~\eqref{SEq:reduced diff eq_a} is solved iteratively using a forward-elimination and back-substitution scheme. We define
\begin{align}\label{Omega defn}
    \Omega_{a,b;n}  \equiv \left(-\na_b^2+i\lambda\right)\psi_{a,b;n}
\end{align}
and solve the recurrence relation:
\begin{align}\label{Eq:ansatz for psistar}
    \psi^*_{a+1,b;n} = E_{a,b}\psi^*_{a,b;n}+F_{a,b;n}
\end{align}
with coefficients determined as:
\begin{align}
&E_{1,b}=-A=i\lambda-2,\,\,\,\,\, E_{a\geq2,b} = -A-(E_{a-1,b})^{-1} \label{SEq:E}\\
&F_{1,b;n}= \Omega_{1,b;n},\,\,\,\,\, F_{a\geq2,b;n}=\Omega_{a,b;n}+\frac{F_{a-1,b;n}} {E_{a-1,b}} \label{SEq:F}
\end{align}
Note that, unlike $F$, $E$ is time independent. These expressions respect the Dirichlet boundary conditions (B.C.):
\begin{subequations}\label{original BC}
    \begin{align}
    &\psi_{0,b;n}=\psi_{\mathbb{J},b;n}=0 \quad \forall \,\,n \label{Seq:BC1,BC2}\\
    &\psi_{a,0;n}=\psi_{a,\mathbb{J};n}=0 \quad \forall \,\,n \label{Seq:BC3,BC4}
    \end{align}
\end{subequations}
i.e., the wave functions vanish at the boundary of the computational box at all times.

The values of  $E_{1,b}$ and $F_{1,b;n}$ are the consequence of the B.C. ~\eqref{Seq:BC1,BC2}. Using the B.C. at $a=\mathbb{J}$, we determine the final point of the intermediate state: 
\begin{align}\label{Eq: psi s J-1}
    \psi^*_{\mathbb{J}-1,b;n} = -F_{\mathbb{J}-1,b;n} \times(E_{\mathbb{J}-1,b})^{-1}
\end{align}
This allows us to reconstruct $\psi^*_{a,b;n}$ across the entire grid at time $n$, first by evaluating for all values of $a \in {0,1,2,..\mathbb{J}}$  at a fixed $b$ and $n$ (which is on the left hand side of Eq.~\eqref{SEq:reduced diff eq_a}) and then repeating the same for all $b \in {0,1,2,...\mathbb{J}}$. Then, $\psi^{*}_{a,b;n}$ is fed to Eq.~\eqref{SEq:reduced diff eq_b}. In order to solve this equation, we define the auxiliary states:
\begin{subequations}\label{Eq:phi and phi^s}
   \begin{align}
    &\phi^*_{a,b;n} \equiv e^{-i\alpha \frac{V}{2}} \psi^*_{a,b;n} \label{SEq:phis and psis}\\
    & \phi_{a,b;n+1} \equiv e^{i\alpha \frac{V}{2}} \psi_{a,b;n+1} \label{SEq:phi and psi}
\end{align} 
\end{subequations}
and solve the second evolution equation as:
\begin{align}\label{Eq:sol phi}
    \left(\na_b^2+i\lambda\right)\phi_{a,b;n+1} = \left(-\na_a^2+i\lambda\right)\phi^*_{a,b;n}
\end{align}
using the same elimination/back-substitution strategy, now applied along the $b$ direction using the B.C. ~\eqref{Seq:BC3,BC4}. The final state $\psi_{a,b;n+1}$ is then constructed via 
Eq.~\eqref{SEq:phi and psi} for all $a,b \in \lbrace 0,1,2..\mathbb{J}\rbrace$.

To summarize the numerical approach, the TDSE for a spin-1/2 particle in two dimensions is solved using the finite-difference method that preserves unitarity. An operator-splitting technique is employed to simplify the time evolution of the state into two decoupled steps by introducing an intermediate wavefunction which is computed iteratively via forward-elimination and back-substitution along one spatial direction, subject to Dirichlet boundary conditions. This intermediate state is then used to propagate the system along the other spatial direction, again using the same iterative strategy. Auxiliary states incorporating the potential term are defined to handle the interaction, and finally, the evolved wavefunction at the next time step is reconstructed across the full grid. This procedure is repeated at each time step, ensuring stable and unitary time evolution throughout the simulation. The flow chart of the numerical approach is illustrated in Fig.~\ref{fig:Flowchart}.

\section{Results}\label{subsec:results}
\subsection{Scattering from 1D skyrmion}\label{subsubsec:1D scattering}
\begin{figure*}[ht] 
    \centering
    \includegraphics[width=0.9\textwidth]{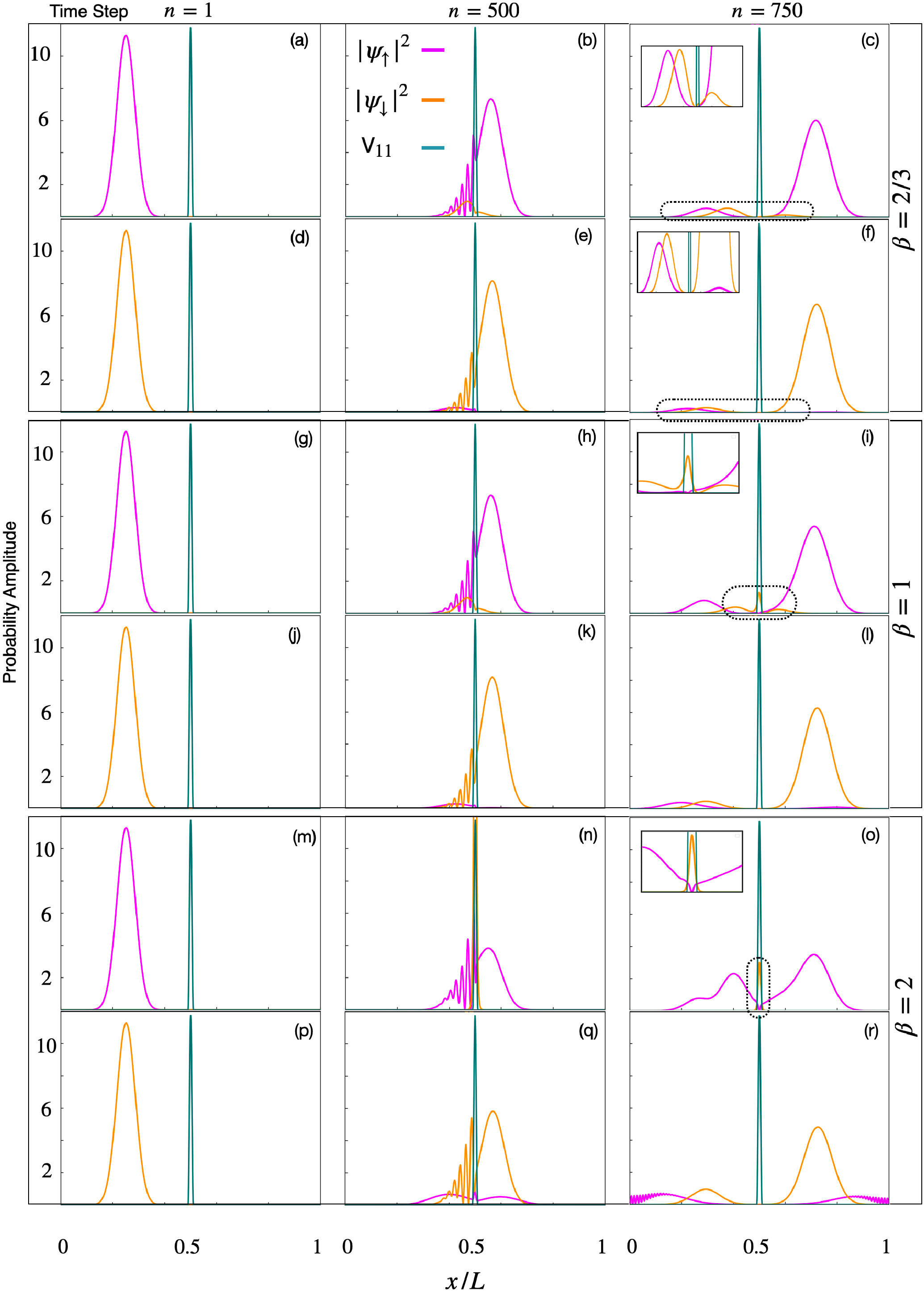}
    \caption{ The snapshots of probability amplitudes at different timesteps($n$) of the scattering of spin-up  (magenta) and spin-down (orange) electron wave packets from a 1D skyrmionic potential as shown in Fig.~\ref{Fig:sky_pot}(a). The potential region is centered at ($0.5L,0.5L$) with a length of $0.04L$. The results are shown for different values of the dimensionless parameter $\beta$ ($ < 1, =1, >1$). The first row in each panel shows the time evolution of the probability density of both spin-up and spin-down electrons with initial spin-up polarization, whereas the second row in each panel represents the same for initial spin-down polarization. The skyrmion region is indicated in teal colored spike. The regions of significance are zoomed in on the insets. In (c) and (f) plots, they highlight the dominance of reflection over transmission in the spin-flip channel when $\beta<1$. In the (i) and (o) plots,  the yellow spikes inside the skyrmion region indicate the presence of the metastable spin-down state.}
    \label{fig:1D_Prob_plot}
\end{figure*}

Since the underlying physics is qualitatively similar, the 1D scenario provides an easier framework for understanding the scattering dynamics and to gain the key insights that will later be extended to the case of scattering from a 2D skyrmion. We note that the numerical scheme described in Section \ref{sec: 2D numerical} is generic and can be straightforwardly implemented for a 1D skyrmion. The spin texture of 1D skyrmion can be obtained from Eq. ~\eqref{Eq:skyrmion coordinate} as follows: 
\begin{align}\label{Eq:1D skyrmion potential}
   &\theta(x) = 
   \left\{
   \begin{array}{ll}
        \left(-\frac{|x|}{a}+1\right)\pi & \text{for } |x| \leq a \\
        0 & \text{for } |x| > a
   \end{array}
   \right.\nonumber\\
   & \phi(x) = \left\{
   \begin{array}{ll}
        \pi & \text{for } -a\leq x < 0 \\
        0 & \text{for } \,\,\,\,\,\,\,0 \leq x < a
   \end{array}
   \right.
\end{align}
The corresponding 1D potential is shown in Fig.~\ref {Fig:sky_pot}(b). A crucial energy scale emerges naturally from Eq.\eqref{Eq:sky pot in}, which is $2J$. 
 
It is important to note that each spin component interacts with two distinct features of the potential during scattering: (i) a scalar potential landscape, which is either a well (for spin-down) or a barrier (for spin-up) of $2J$, and (ii) a spin-flip term arising from the off-diagonal elements of the potential matrix (Eq.\eqref{Eq:sky pot in}). The scattering dynamics are highly dependent on the dimensionless parameter $\beta=\frac{2J}{\text{KE}}$ and exhibit qualitatively different behavior depending on its value, i.e., $\beta<1, \beta=1,\beta>1,$ due to the interplay between potential depth/barrier and spin-flip effects. The results are summarized through Fig~\ref{fig:1D_Prob_plot}.

\noindent
\textit{Case 1- $\beta<1$ High kinetic energy}: When the incident kinetic energy exceeds the critical energy scale $2J$, the spin-preserving component transmits through the potential region with minimal reflection (see  Figs.~\ref{fig:1D_Prob_plot}-(c) and \ref{fig:1D_Prob_plot}-(f) ). This minor reflection occurs due to the low-momentum modes present in the Gaussian tail. However, the spin-flip components, which are dynamically generated inside the skyrmion, despite having large kinetic energy, show higher reflection than transmission (see the inset plot in  Figs.~\ref{fig:1D_Prob_plot}-(c) and \ref{fig:1D_Prob_plot}-(f) ). This non-trivial phenomenon is due to the occurrence of a destructive interference between the forward and backward moving spin-flip components in the second half of the potential region. A detailed explanation of this destructive interference is provided in the Appendix \ref{Appendix-2} material.

\noindent
\textit{Case 2- $\beta=1$ Intermediate kinetic energy}: When the kinetic energy is comparable to $2J$, the spin-preserving components show similar behavior to the case of  $\beta<1$, but now the probabilities of transmission and reflection are lower and higher, respectively (see  Figs.~\ref{fig:1D_Prob_plot}-(i) and \ref{fig:1D_Prob_plot}-(l)). Apart from the non-trivial reflection in the spin-flip channel, interestingly, a meta-stable spin-down state is dynamically generated from the spin-up state via flipping  (see the little kink inside the skyrmion region in the of Fig.~\ref{fig:1D_Prob_plot}-(i)). However, over a long period of time, this meta-stable state escapes the skyrmion region by flipping again to the spin-up state, and its explanation is discussed in the next case. 

\noindent
\textit{Case 3- $\beta>1$ Low kinetic energy}: As expected, there is a further decrease in transmission and an increase in reflection by lowering the kinetic energy. Most importantly, with an increase in $\beta$ the transmission probability asymptotically decreases to a saturation. The underlying mechanism will be explained in the next paragraph. Furthermore, the metastable state is more prominent in this case. The double bump in the probability density of the spin-up electron and the kink in the spin-down electron (see Fig.~\ref{fig:1D_Prob_plot}-(o) ) emphasize our earlier explanation of the trapped spin-down state moving out as the spin-up state. This statement is corroborated by artificially disabling the $\Vud$ matrix element (not presented in the plots), which is responsible for down-to-up conversion, in which the spin-down state gets trapped inside the scatterer for all time. Further, in all cases of scattering, the spin-up electrons always move faster than their spin-down counterparts, which has been verified numerically by computing the expectation value of the kinetic energy as discussed in the Appendix \ref{Appendix-1} material.

\begin{figure}
    \centering
    \includegraphics[width=1\linewidth]{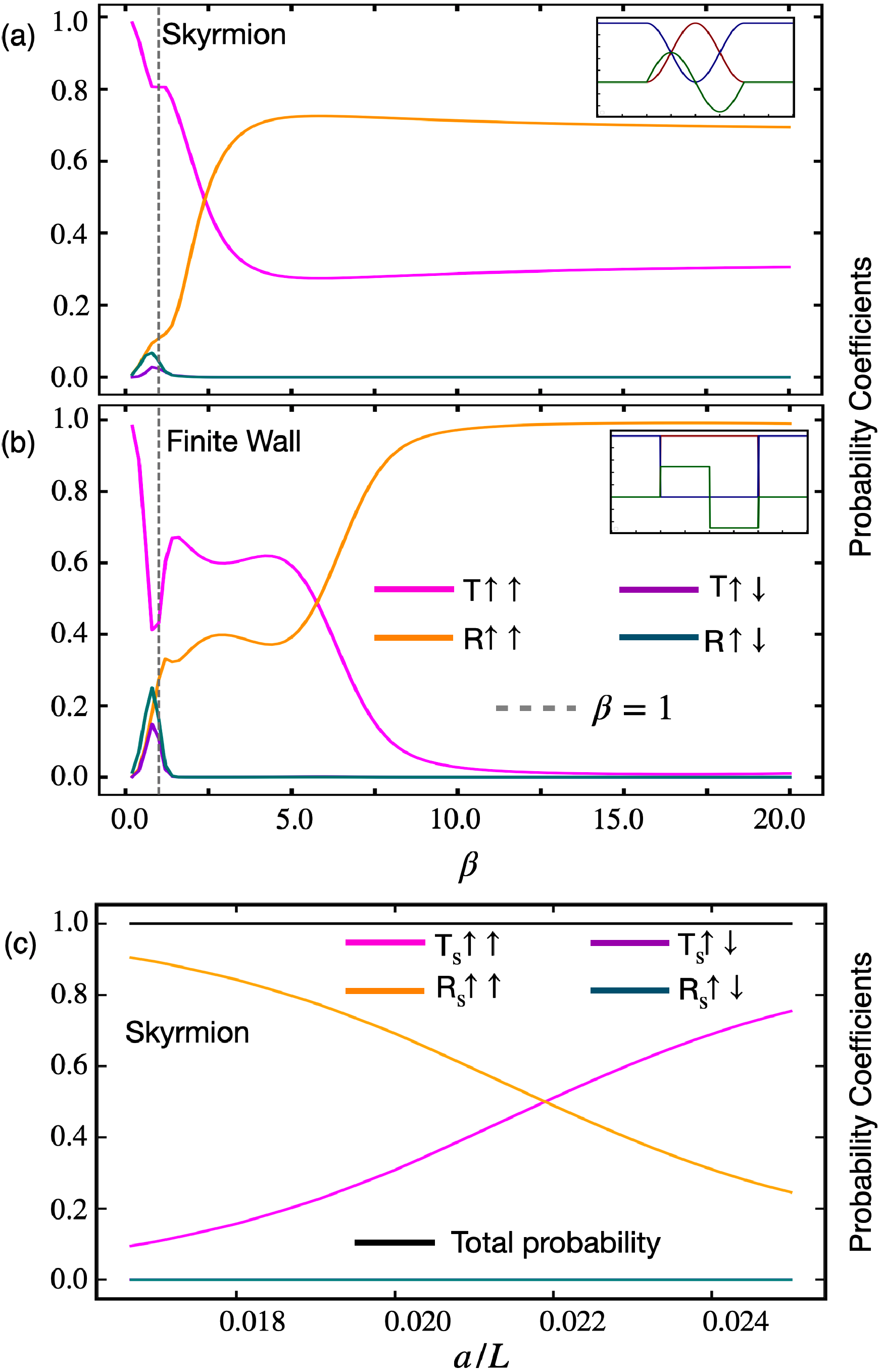}
    \caption{Different plots of the transmission coefficient ($\mathrm{T}$) and the reflection coefficient ($\mathrm{R}$) of $\uu$ and $\ud$ channels between a skyrmion potential (a) and a step potential (with all $\Vuu$, $\Vdd$, $\Vud$ and $\Vdu$ non-zero values potential) (b) and the plots show the variation of $\mathrm{T}$  and $\mathrm{R}$ as a function of $\beta$. (c) The saturation value of the different transmission and reflection coefficients ($\mathrm{T}_s$  and $\mathrm{R}_s$)is plotted as a function of skyrmion size $a$, which is in the units of $L$.
    }
     \label{fig:T and R saturation}
\end{figure}

In Fig.~\ref{fig:T and R saturation}(a), we quantify the transmission and reflection, both for spin-preserved and spin-flip channels, as a function of $\beta$. The salient feature we get is that, in the limit $\beta \gg 1$, the reflection and transmission probabilities for the $\uu$ channel (spin up electron scattered as spin up electron) saturate at finite values while the $\ud$ (spin up electron scattered as spin down electron)probability vanishes. This finite saturation in the $\uu$ channel is counterintuitive, as one expects the spin-up component to be almost completely reflected at large $\beta$. Two mechanisms could account for this behavior. Firstly, in the absence of a spin-flip potential, the problem reduces to the scattering of a spinless particle from a potential barrier, where the transmission probability asymptotically vanishes with a large barrier. Therefore, spin-flip is necessary to have a finite transmission probability. Secondly, there is a role of non-collinearity in spin texture. To disentangle it, in a hypothetical design, we suppress the non-collinearity while retaining the spin-flip coupling by adopting the following simplified model:
\begin{align}\label{Eq:Model spin flip}
    &\theta(x) = 
   \left\{
   \begin{array}{ll}
        \pi & \text{for } |x| \leq a \\
        0 & \text{for } |x| > a
   \end{array}
   \right.\nonumber\\
   & \phi(x) = \left\{
   \begin{array}{ll}
        \pi & \text{for } -a\leq x < 0 \\
        0 & \text{for } \,\,\,\,\,\,\,0 \leq x < a
   \end{array}
   \right.  
\end{align}     
which renders the components of the potential matrix to vary as a piecewise step function (refer to the inset plot of Fig.~\ref{fig:T and R saturation}(b)). In the absence of non-collinearity, in the large $\beta$ limit, the reflection and transmission probability saturate to one and zero, respectively. This makes the non-collinearity a sufficient condition to achieve a finite transmission probability. Interestingly, this saturation of the transmission is related to the skyrmion size. In the $\beta \gg1$ regime, as the skyrmion size increases, the electron spin aligns smoothly along the local magnetization direction, resulting in less reflection and more transmission.

Taken together, the 1D scattering analysis reveals distinct regimes of spin-resolved transport. For spin-preserving channels, the reflection and transmission probabilities remain trivial in the weak-interaction limit ($\beta \ll 1$), but develop a non-trivial saturation behavior at strong-interaction ($\beta \gg 1$). In contrast, the spin-flip channels exhibit a significantly higher reflection than transmission even in the $\beta < 1$ regime— an unexpected feature absent in the spin-preserving sector. Interestingly, the spin-flip dynamics reveal the emergence of a metastable spin-down state inside the skyrmion, whose relaxation proceeds through an iterative flipping mechanism that becomes prominent once $\beta \geq 1$.

\subsection{The 2D scattering}\label{subsubsec:2D scattering}
\begin{figure}
    \centering
    \includegraphics[width=0.9\linewidth]{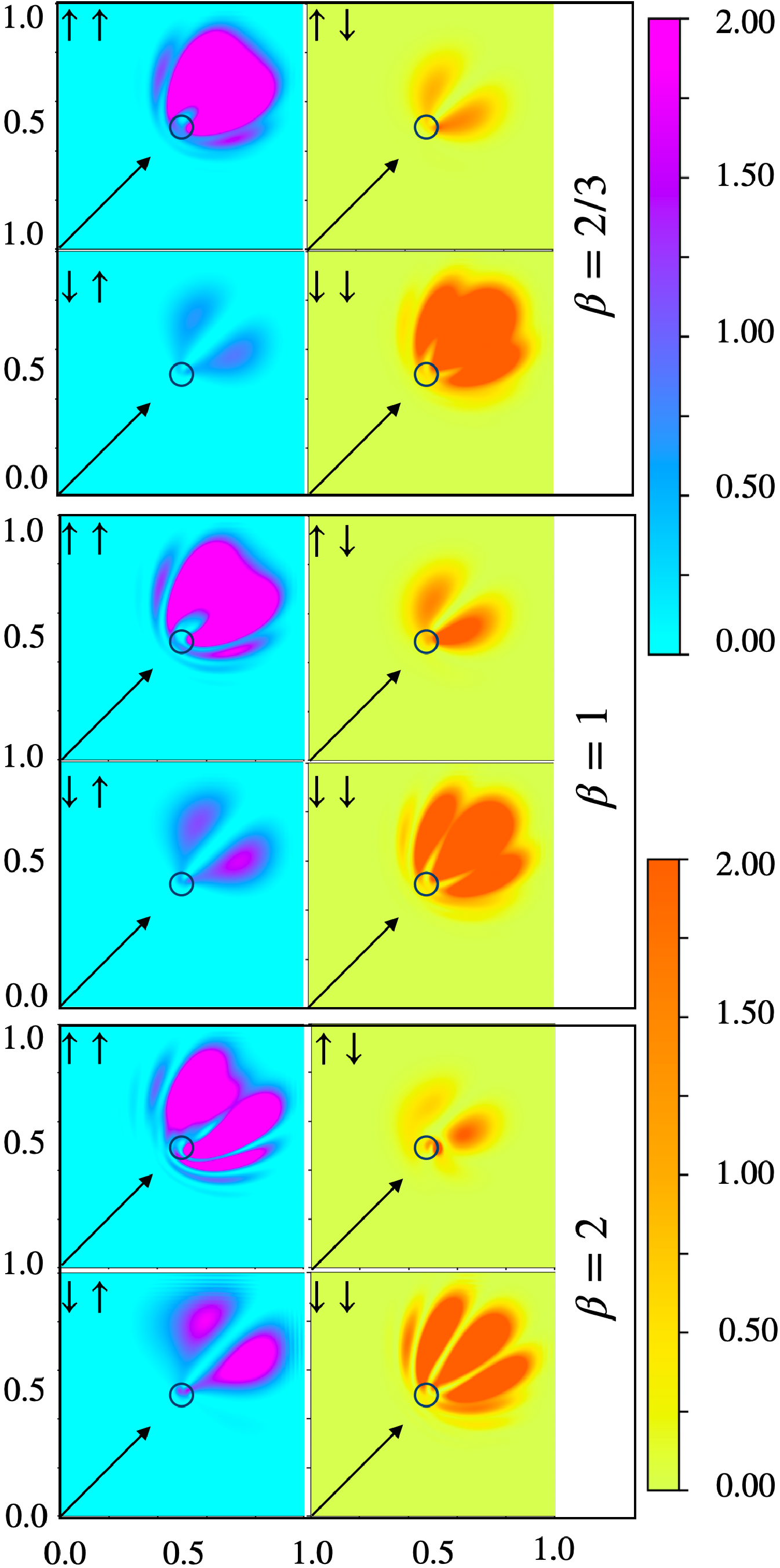}
    \caption{ Snapshot of time evolution of the electron wave function after scattering from a skyrmion for different $\beta=\frac{2J}{\text{KE}}$. The real-time dynamics is provided in supplementary material \cite{supplementary} This represents the 2D counterpart of 1D results shown in the third column of Fig.~\ref{fig:1D_Prob_plot}. The skyrmion region is marked by the black circle. The $45^{\circ}$ arrow shows the direction of initial propagation. The blue-magenta gradient represents the probability distribution of the spin-up particle, whereas the yellow-orange gradient represents the same for spin-down. The maximum of the color gradients is set at 2.0 for a clear visualization of the scattered state with lower probability.}
    \label{fig:2D-Full-Scattering}
\end{figure}

Building on the key features observed in the one-dimensional scattering scenario, in this section, we extend our analysis to scattering of electrons in a 2D skyrmion, the potential for which is shown in Fig.~\ref{Fig:sky_pot}(a). While the 1D model offers meaningful insight into the core mechanisms of the scattering process, the transition to 2D brings additional insights. In the 1D case, all parts of the Gaussian wave packet are subject to the same potential profile, as shown in Fig.~\ref{Fig:sky_pot}(c). However, in two dimensions, the presence of an azimuthal degree of freedom, $\phi$, introduces spatial anisotropy. Although different regions of the wave packet encounter qualitatively similar features of the potential, the effective height of the potential barrier varies with $\phi$, resulting in direction-dependent scattering outcomes that are absent in the 1D case. The results corresponding to scattering from 2D skyrmion for different energy regimes are shown in Fig.~\ref{fig:2D-Full-Scattering}.

Irrespective of the value of $\beta$ and initial spin polarization, the large fraction of spin-preserving channels moves along the forward direction, and a small portion backscatters. However, the fraction of back scattering increases with an increase in $\beta$ as expected.  This can be inferred from the higher number of wrinkles in $\uu$/$\dd$(spin down electron scattered as spin down electron) channel as we increase $\beta$ (see Fig.~\ref{fig:2D-Full-Scattering}). We may note that some back scattering at low $\beta$ appears due to the infrared tails of the Gaussian distribution and is less prominent in the figure.

Unlike the spin-preserving channels (see Fig.~\ref{fig:2D-Full-Scattering}), the spin-flip channels are highly dependent on $\beta$ following Eq.~\eqref{Eq:sky pot in}. The scattering intensity of the $\du$(spin down electron scattered as spin up electron) channel along the forward direction increases with $\beta$. However, it is not the case in the $\ud$ channel. The reason is explained as follows. Even though the probability of $\ud$ flipping is the same as in the case of $\du$ (see Eq. ~\eqref{Eq:sky pot in}),  the dynamically generated spin-down electrons inside the skyrmion via spin-flipping have to overcome the $2J$ barrier in order to propagate outside of the scatterer. Therefore, regardless of the fact that $\ud$ conversion increases with $\beta$, due to a lack of sufficient kinetic energy, the spin-down states stay inside, except for some quantum tunneling, as can be seen in the $\ud$ plot in the lower panel of Fig.~\ref{fig:2D-Full-Scattering}. Nevertheless, these trapped spin-down states escape the barrier by back flipping to the spin-up state. This phenomenon can be observed from secondary waves whose elongated tails are present inside the skyrmion (see the $\uu$ channels in Fig.~\ref{fig:2D-Full-Scattering}). Furthermore, the presence of the angular phase factor $e^{\pm{i\phi}}$ introduces a spin to orbital momentum transfer in the spin-flip channel dynamically. This occurs to conserve the total angular momentum ($-i\partial_{\alpha}+m \hat{s}_z$). We may note that real-time spin-angular momentum transfer has not been reported yet.

In the scattered state, the expectation value of the kinetic energy of the spin-up electron is always higher than that of the spin-down state, as the former slides downhill and the latter has to climb uphill, as shown in Fig.~\ref{Fig:sky_pot}, and the same has been validated numerically as discussed in the Appendix \ref{Appendix-1} material.

\subsection{Scattering cross section}
\begin{figure*}[ht] 
    \centering
    \includegraphics[width=\textwidth]{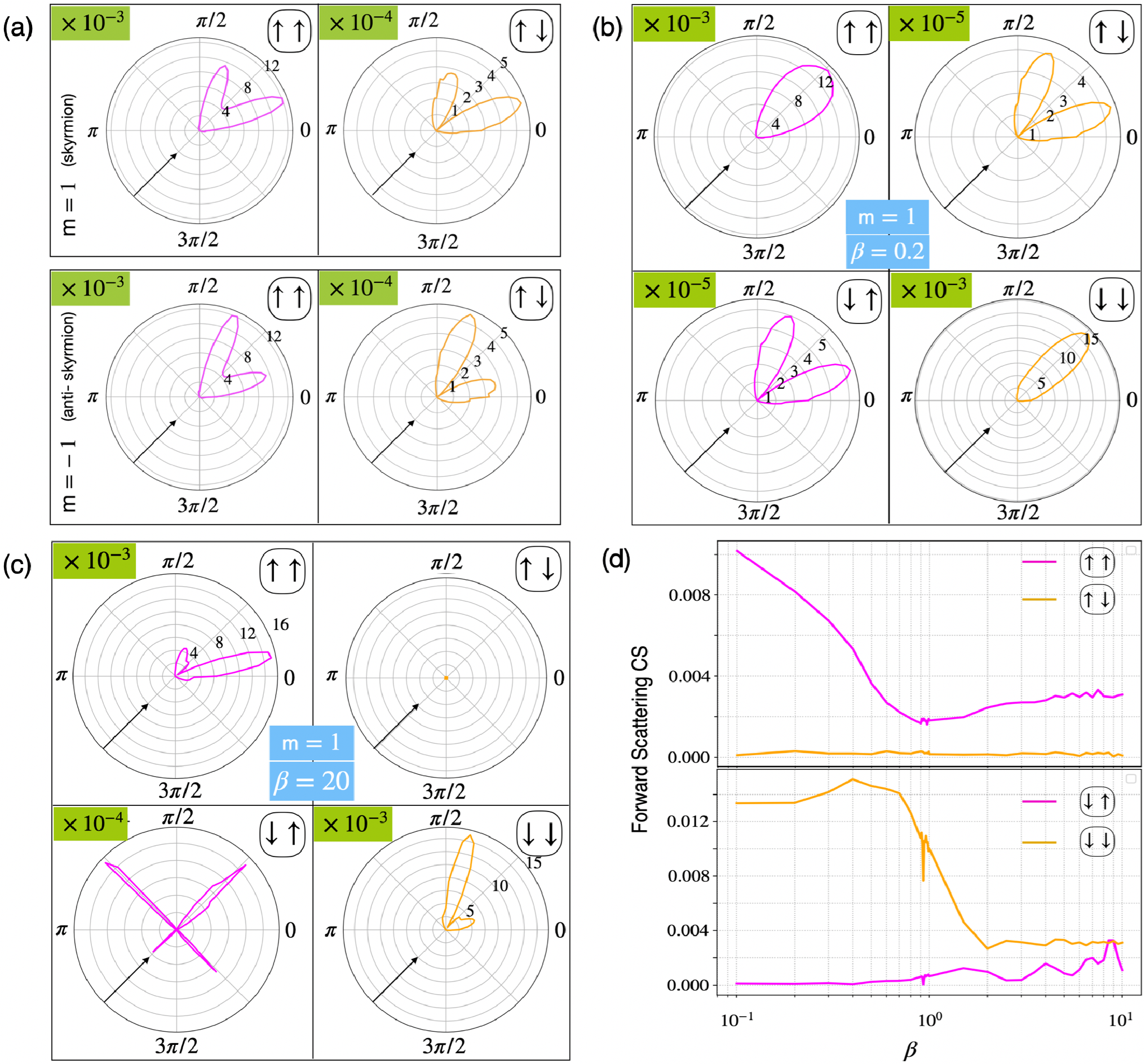}
    \caption{The plots of the scattering cross-section for different helicities and interaction strengths. The $45^{\circ}$ arrow shows the direction of initial propagation. (a)The plot of the scattering cross-section for chirality $m=1$ and $m=-1$ reveals the anti-symmetric nature of the quantity for a skyrmion and an anti-skyrmion. (b)The plot of scattering cross-section at low interaction strength $\beta=0.2$. At this energy scale, all the scattering channels exhibit symmetry along the propagation axis. (c) The plot of scattering cross-section at high interaction strength $\beta=20$, where the symmetry of the spin-preserving channel is lost with spin-up and spin-down scatter in opposite directions. The spin-flip channels are almost suppressed. (d) Plot of forward scattering as a function of coupling strength. The $ x-$axis is in the $\log$ scale. The forward scattering for the spin-preserving channel decreases with an increase in the coupling strength and saturates close to zero. }
    \label{fig:2D_scattering CS}
\end{figure*}

The scattering cross section is the fundamental observable that classifies particle interactions and bridges theoretical predictions and experimental measurements. In the present study, we compute the differential scattering cross section along a particular direction by computing the fraction of propagating modes present along that direction. In addition to the energy-dependent behavior, the two-dimensional framework enables an exploration of the roles played by helicity ($\gamma$) and vorticity ($m$) of the skyrmion spin texture in the scattering process. The scattering cross section for various conditions is shown in Fig.~\ref{fig:2D_scattering CS}. The important inference we make here is that the scattering cross section remains invariant with $\gamma$, indicating robustness of the scattering profile against changes in the spin texture's winding. In contrast, the scattering exhibits antisymmetry under vorticity reversal ($m \rightarrow -m$) as shown in Fig.~\ref{fig:2D_scattering CS}(a), revealing a directional asymmetry inherently induced by the underlying skyrmion texture.

In the weak interaction regime ($\beta \ll 1$) as plotted in Fig.~\ref{fig:2D_scattering CS} (b), all scattering channels exhibit mirror symmetry with respect to the incident propagation axis. The spin-preserving channels ($\uu$, $\dd$) display maximal differential cross sections along the forward (incident) direction, whereas the spin-flip channels ($\ud$, $\du$) vanish along this axis due to destructive interference as already discussed for the 1D case. On the contrary, as $\beta$ increases, this symmetry is progressively broken. As plotted in Fig.~\ref{fig:2D_scattering CS}(c), in the strong interaction regime ($\beta\gg1$), the spin-resolved scattering becomes highly asymmetric: spin-up and spin-down components are preferentially scattered into opposite transverse directions. The $\ud$ channel is strongly suppressed and effectively vanishes, while the $\du$ channel remains finite even though it is suppressed by approximately an order of magnitude relative to the dominant spin-preserving channels. Further, the scattering cross section of the spin-preserving channel saturates at high interaction strength ($\beta$) with transmission probability asymptotically vanishing (see Fig.~\ref{fig:2D_scattering CS} (c) and (d)).

\section{Summary and Outlook}

\subsection{Summary}
In this present work, the time evolution of a spin-polarized Gaussian wave packet of electrons has been studied by explicitly solving the TDSE numerically. The finite difference method is followed in our numerics and an iterative propagation scheme has been implemented based on forward elimination and back substitution, along with the operator splitting technique. Our numerical scheme has second order convergence both in space and time as shown in Appendix \ref{Appendix-3}.  Using the designed numerical framework, we have studied the scattering dynamics of the electron wave packets interacting with chiral spin textures of skyrmions in both 1D and 2D geometries. Our simulations have revealed that the scattering process is insensitive to the chirality, $\gamma$. However, in agreement with the prior observations \cite{Denisov_2017}, the differential scattering cross section for skyrmion and anti-skyrmion (skyrmion with opposite vorticity) exhibits a mirror symmetry along the wave packet's propagation axis. A key advantage of our approach lies in its ability to capture the full spatio-temporal evolution of the wave packet, thereby providing physical insights into how the competition between the kinetic energy (KE) and skyrmion potential arising from the Hund's coupling ($J$) between itinerant electrons and the local spin controls the spin-dependent scattering dynamics. Consequently, an important parameter $\beta (=\frac{2J}{KE})$ emerged to describe the scattering phenomena in this quantum system.

The key observations are as follows: (i) The electron iteratively flip their spin within the skyrmionic region and significantly influence the outcome. (ii) In the spin-flip scattering channels (e.g., an incoming spin-up wave giving rise to a reflected and transmitted spin-down wave), the transmission probability remains significantly suppressed compared to reflection, even in the low-$\beta$ regime. This occurs as the flip potential is composed of two equal and opposite halves inside the skyrmion (see Fig.~\ref{Fig:sky_pot}). While the first half determines the reflection probability, the second half influences the transmission. The forward-moving waves emerging from the first half experience destructive interference with the backward-moving wave from the second half, which results in a lowering of transmission probability. (iii) Our simulations both reveal distinct secondary wavefronts associated with back-converted spin components—features that, to our knowledge, have not been reported previously. (iv) We observe the formation of long-lived quasi-bound states localized near the scattering center, which persist until disrupted by boundary-induced reflections. (v) Concerning the scattering in spin-preserved channel, counterintuitively, we find that in the high-$\beta$ regime where the interacting potential strength is high,  both transmission and reflection probabilities remain finite. Furthermore, the scattering cross-section loses its symmetry with increasing $\beta$, and the forward scattering (along the propagation axis) asymptotically vanishes as $\beta$ increases.


In a very recent work \cite{skyrmion-scattering}, experimenters have been successful in employing pump-probe X-ray microscopy to study the current-driven dynamics of mobile skyrmions from the stationary skyrmions at the nanosecond level. Therefore, the present study of electron scattering from a stationary skyrmion is possible to validate using an identical or a similar methodology. The experimental realization of the intermediate states and iterative spin-flipping will open new avenues to further explore the skyrmion dynamics.

\subsection{Outlook} 
The theoretical framework established here is generic to arbitrary NCS such as bimeron, skyrmionium. Therefore, it has the potential to explore many exotic quantum dynamical phenomena. Furthermore, the dynamical richness can be explored in periodic NCS arrays forming crystals which exhibit non-trivial quantum transport. A promising future direction is to extend the TDSE framework to capture the full complexity of electron–skyrmion quantum dynamics by incorporating Rashba or Dresselhaus spin–orbit coupling (SOC). This SOC, often generated either through external field or through substrate engineering, redistributes the emergent magnetic field and effective potential and thereby the scattering dynamics \cite{mandal, Rashba-Dresselhaus-2016, rb-elec}. In combination with skyrmion-induced emergent fields, SOC has the ability to tailor spin–orbit entanglement \cite{blugel}, to enable asymmetric scattering, and to facilitate ultrafast, electric-field–controlled spin filtering, pumping, or trapping. 
Importantly, the TDSE framework directly yields entanglement entropy, providing a quantitative measure of decoherence, spin–orbital entanglement, and information flow, thus linking skyrmion-based transport to quantum information science and paving the way for topologically robust, electrically tunable spintronic and quantum computing platforms\cite{Yang_2015,Balents2010,qubit1}.

\section*{Acknowledgments} 
AM thanks MoE India for the PMRF fellowship.  SS thanks SERB India for the VAJRA fellowship. The authors acknowledge the valuable input of Saurav Samantaray while formulating the numerical method.



\section*{Data Availibility}
The data that support the findings of this article are not publicly available. The data are available from the authors upon reasonable request.

\appendix

\section{Self-Adjointness of a quantum operator}\label{Appendix-1}

\noindent
Consider an operator $\mathcal{L}$ such that 
\begin{align}\label{App:eq:defn of L}
    \mathcal{L}: \mathcal{V} \rightarrow \mathcal{V}
\end{align}
where, $\mathcal{V}$ is the vector space and $\phi,\,\,\psi \in \mathcal{V}$. The inner product on $\mathcal{V}$ is defined as
\begin{align}\label{App:eq:inner product defn}
    \braket{\phi|\psi}=\int_a^b \phi^{*}(x) \psi(x) dx
\end{align}

Then, $\mathcal{L}^{\dagger}$, the adjoint of $\mathcal{L} $  is defined as 
\begin{align}\label{App:eq:defn adjoint}
    \braket{\phi|\mathcal{L\psi}}\equiv\braket{\mathcal{L}^{\dagger}\phi|\psi}
\end{align}

If $\mathcal{L} = \mathcal{L}^{\dagger}$ then the operator is self-adjoint and if $\mathcal{L} = -\mathcal{L}^{\dagger}$ then it is anti-self-adjoint. Note, that the boundary conditions play a crucial role for the (anti-) self-adjointness of an operator, which is mostly ignored or assumed. However, during numerical simulations, if this is not addressed properly, one tends to pick up non-real expectation values due to the loss of (anti-)self-adjointness of the operator.

Here, we will show how to deal with it by taking the example of the momentum operator. In the unit of $\hbar=1$
\begin{align}\label{App:eq:defn momentum operator}
    \mathcal{L}_p=-i\frac{d}{dx}
\end{align}
Therefore,
\begin{align}
     \braket{\phi|\mathcal{L}_p\psi} & = \int_a^b dx\,\, \phi^{*}(x)\left(-i\frac{d}{dx}\right)\psi(x) \nonumber\\
                                   & = \left(\int_a^b dx\,\,  i \frac{d}{dx}\left(\phi^*(x)\right)\psi(x)\right) - i \phi^*(x)\psi(x)\Bigg|_a^b
\end{align}

Now, $\mathcal{L}_p$ will be self-adjoint provided the second term on the right-hand side vanishes i.e.

\begin{align}\label{App:Eq:bdy condn on self adjoint}
    \phi^*(b)\psi(b)-\phi^*(a)\psi(a)=0
\end{align}
The following sets of boundary conditions make sure the momentum operator in Eq.~\eqref{App:eq:defn momentum operator} is self adjoint:

\begin{enumerate}
    \item $\psi(a) = \psi(b) = 0$. Equivalently, $\psi^{(n)}(a) = \psi^{(n)}(b) = 0$ where, $\psi^{(n)}(x)$ is the $n-$th derivative at $x$.
    \item $\psi(a) = \beta \psi(b)$ with $|\beta| =1$.
\end{enumerate}

\vspace{1cm}

Therefore, the self-adjointness of an operator relies on applying correct boundary conditions. While dealing with quantum mechanics, this thing is automatically taken care of when we apply the postulates of the theory to the wave functions; however, the problem of non-self-adjointness arises when one tries to compute physical observables like expectation values of the operator for an interval that differs from the original boundaries of the problem. In such cases, during numerical computations, one tends to pick up the non-trivial imaginary component on the right-hand side of  Eq.~\eqref{App:Eq:bdy condn on self adjoint}. Therefore, a redefinition of the operator is required while writing the code,e.g., if we change the boundary from $(a,b)$ to $(\tilde{a},\tilde{b})$, then we can define the expectation value of the momentum operator as
\begin{align}\label{App:Eq:redefinition_of_momentum_operator}
    \braket{\hat{p}} = \frac{1}{2} \Bigg(& \int_{\tilde{a}}^{\tilde{b}} dx\,\, \psi^*(x) \left(-i\frac{d\psi(x)}{dx}\right) \nonumber \\
    & + \int_{\tilde{a}}^{\tilde{b}} dx\,\, \left(i\frac{d\psi^*(x)}{dx}\right)\psi(x) \Bigg)
\end{align}
to remove the imaginary part that pops up during numerics.

\section{The near-zero transmission in the spin-flip channel}\label{Appendix-2}

\begin{figure}[h]
    \centering
    \includegraphics[width=1\linewidth]{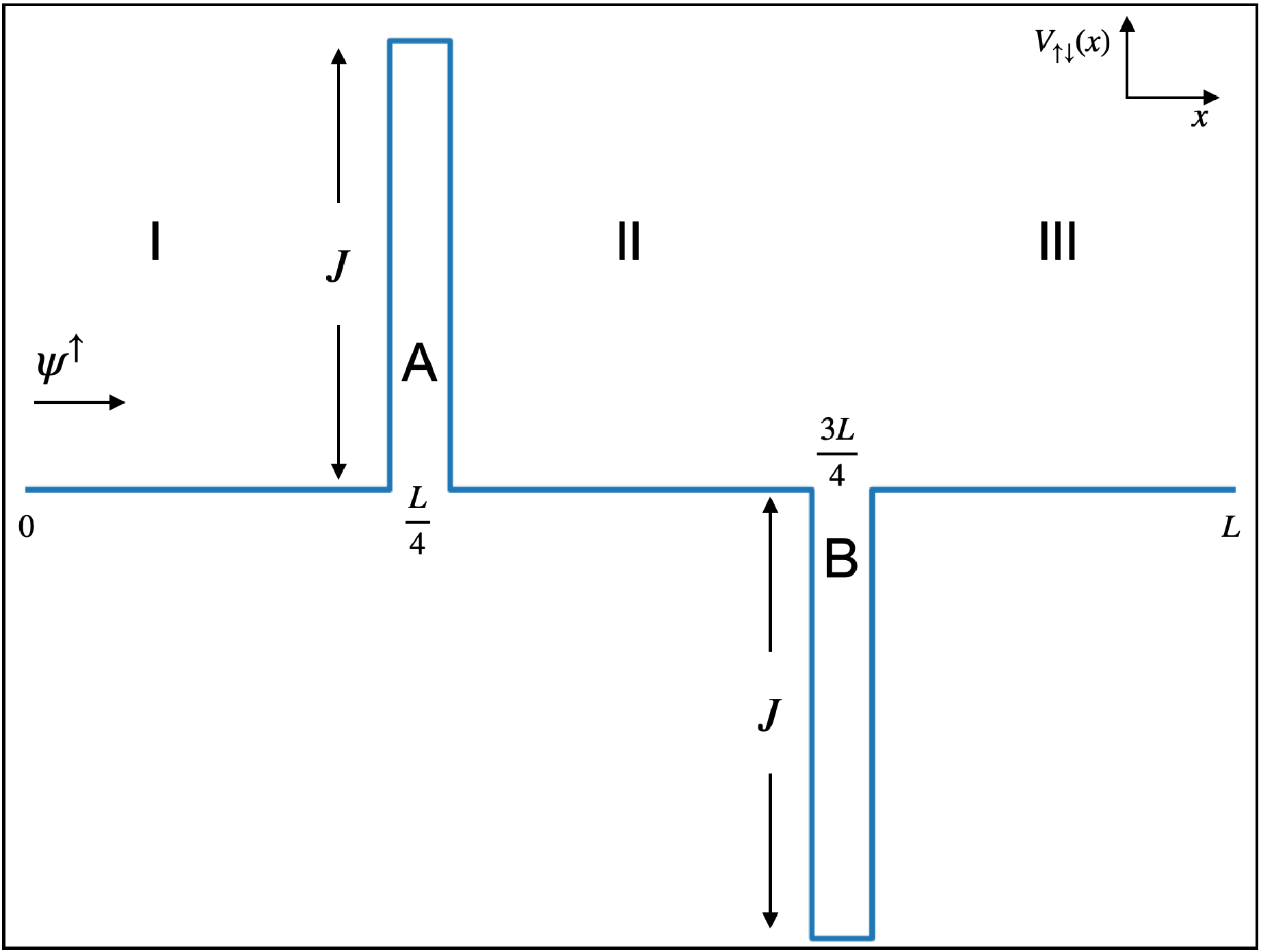}
    \caption{The plot of the off-diagonal term of the model potential defined in Eq.~\eqref{Eq:model potantential for near zero transmission}.  }
    \label{fig:near_zero_trnas}
\end{figure}

In the weak interaction limit i.e., low $J$, we have reported the following observations in the spin-flip channels:
\begin{enumerate} [label=(\roman*)]
    \item the transmission is suppressed substantially in comparison with reflection in the case of a 1D skyrmion.
    \item the corresponding 2D behavior reflects, the scattering amplitude is zero along the incident direction.
\end{enumerate}
Here, we are going to  analytically argue the same  by considering a model potential of the following  kind
\begin{align}\label{Eq:model potantential for near zero transmission}
    V(x)  = 
   \left\{
   \begin{array}{ll}
        \begin{pmatrix}
     0 & 0\\
     0 & 2J
    \end{pmatrix}  & \text{in region I, II and III }  \\
    \begin{pmatrix}
     2J & J\\
     J & 0
    \end{pmatrix}  & \text{in region A }  \\
    \begin{pmatrix}
     2J & -J\\
     -J & 0
    \end{pmatrix}  & \text{in region B } 
   \end{array}
   \right.\nonumber\\
\end{align}
and in Fig.~\ref{fig:near_zero_trnas} we have represented the potential region only with $\Vud$ plot; however, other components can be understood from that reference.

Let's consider that we start with a spin-up state. Then we need to show that the probability of the spin-down state in region III is smaller than that in region I. Following Eq.~\eqref{Eq:schrodinger equation for spin 1/2 position basis} 
\begin{align}\label{App:Eq: spin down state}
    \psid \propto \Vdu \psiu + \Vdd \psid
\end{align}
and we summarize the wave function in different regions at different times in ~\ref{Table: near-zero-trans_table} where

\begin{align*}
    &t_1:\, \text{time before the interaction}\\
    &t_2: \text{ time during the interaction}\\
    &t_3: \text{ time after the interaction}
\end{align*}

\begin{table}[h!]
\centering
\begin{tabular}{|c|c|c|c|}
\hline
Time & I & II & III \\ \hline
$t_1$ & 0 & 0 & 0 \\ \hline
$t_2$ & 0 & $J \psiu$ & 0 \\ \hline
$t_3$ & $(2J^2+J)\psiu$ & 0 & $(2J^2-J)\psiu$ \\ \hline
\end{tabular}
\caption{ $\psid$ at different time intervals.}
\label{Table: near-zero-trans_table}
\end{table}
Now, considering the low interaction strength, the probability of the dynamically generated spin-down state after all the interactions have happened
\begin{align}
  |\psid|^2 \propto \left\{ 
  \begin{array}{ll}
  J^2+2J^3+\mathcal{O}(J^4) \quad \text{in region I }\\
  J^2-2J^3+\mathcal{O}(J^4)   \quad \text{in region III }
  \end{array}   
  \right.
\end{align}
Therefore, the reflection dominates over transmission in the sub-leading order.

\begin{figure}[hbt!]
    \centering
    \includegraphics[width=0.9\linewidth]{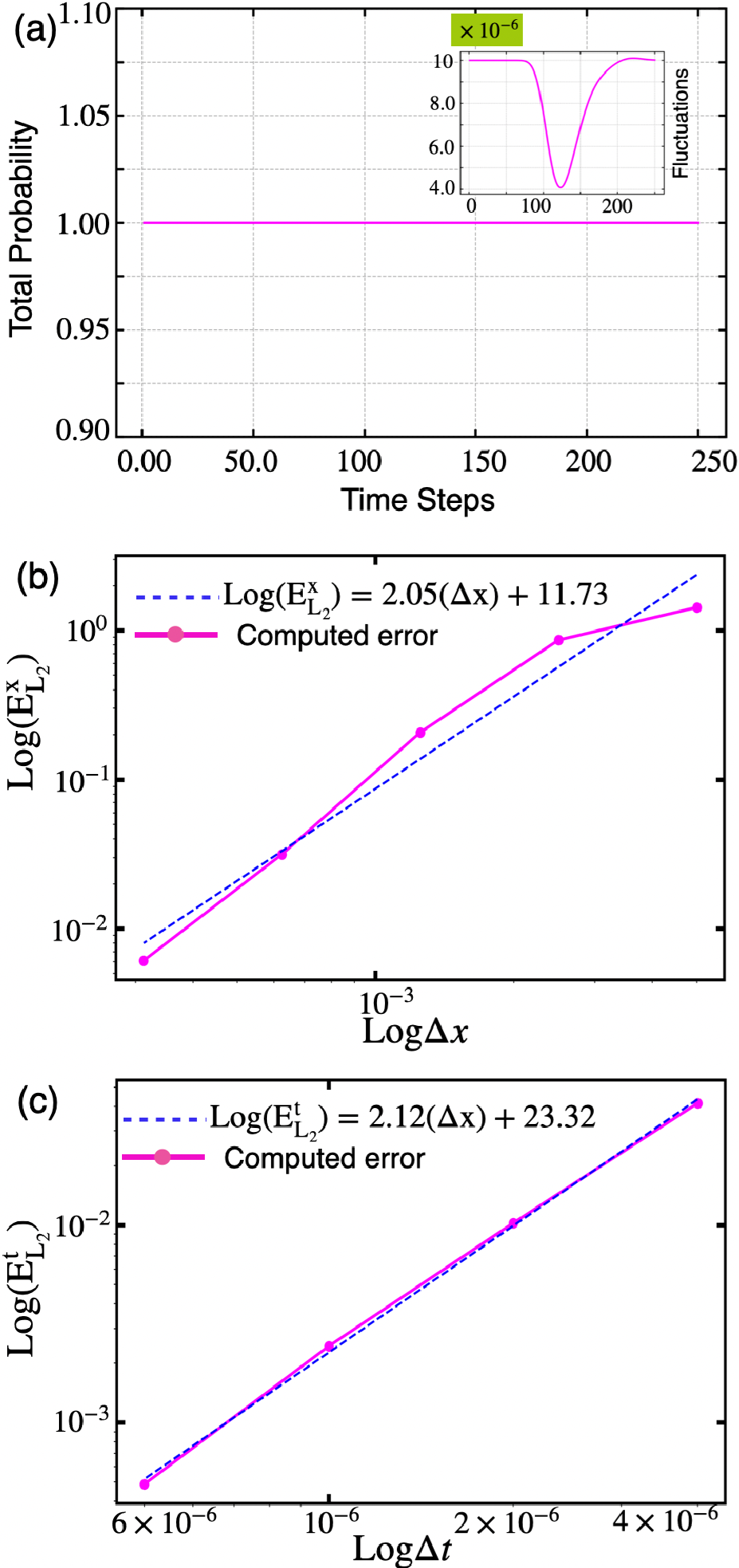}
    \caption{(a)The plot of the total probability as a function of time steps, and the inset plot shows the fluctuations. This indicates that our numerical scheme conserves the total probability to within an error of $10^{-6}$. (b) The log-log plot of the $E_{L_2}^x$ vs. spatial step size $\Delta x$. Magenta dots represent the computed errors, and the blue dashed line indicates the linear fit. (c) The log-log plot of the $E_{L_2}^t$ vs. temporal step size $\Delta t$. Magenta dots represent the computed errors, and the blue dashed line indicates the linear fit.}
    \label{fig:Probability_conservation test}
\end{figure}

\section{Convergence test and error analysis}\label{Appendix-3}
In order to comment on the accuracy of the numerical scheme, we plot the total probability as a function of time in Fig. \ref{fig:Probability_conservation test}(a), which is conserved to unity with accuracy up to the sixth decimal point. Note that this is a consequence of the splitting of the time evolution operator in Eq. \ref{Eq: unitary e-i alpha H} to preserve the unitarity. For further validation, we perform a convergence test on both the spatial and temporal grids by computing the  $L_2$ (square integrable) error norm. Let us consider that  $F(x_{i_{\rm{ref}}})$ is the reference function on the defined finest mesh for $X$ where $x_{i_{\rm{ref}}}\in X $ are the reference mesh points. Now if the coarse function $G$ available on the coarse mesh points $x_{i_{\rm{c}}}\in X$, then the $L_2$ error norm is  defined as 
\begin{align}\label{L2 error norm definiton}
 E_{L_2}^x = \sqrt{\sum_{i} \bigg| F(x_{i_{\rm{ref}}}=x_{i_c})-G(x_{i_{c}})\bigg|^2 \Delta x}  
\end{align}
where $\Delta x$ is the uniform spatial step size on the coarse mesh. If 
\begin{align}\label{Eq:order of converegence defn}
    E_{L_2} \sim (\Delta x)^n 
\end{align}
then, the spatial order of convergence is said to be $n$ \cite{Nwaigwe2025,Saurav_2024}. Below, we illustrate the convergence test for the case of the electron scattering from the 1D skyrmion. The analysis remains valid for the other cases presented here. The reference finest mesh is designed with  total number of mesh points $N_{\rm{ref}}=6400$.

To determine the spatial order of convergence for our numerical scheme, we fix the time step size $\Delta t$ and the coarse meshes are taken to be $N_c=\frac{1}{\Delta x}=\lbrace{200,400,800,1600\rbrace}$. The log-log plot of $E_{L_2}$ vs. $\Delta x$ is plotted in Fig. \ref{fig:Probability_conservation test}(b). The slope is found to be $\sim 2.05$, determining the spatial order of convergence is $2$. This is because the spatial discretization of our differential operator

\begin{align}\label{Eq:second order discretization}
    \frac{\partial^2\psi}{\partial x^2} \approx \frac{\psi_{i+1}-2\psi_i+\psi_{i-1}}{\Delta x^2} +\mathcal{O}(\Delta x^2)
\end{align}
carries a leading order error term proportional to $\Delta x^2$.

Next, we compute the temporal order of convergence by fixing the spatial step size $\Delta x$ and varying the time step size $\Delta t$ by varying $\lambda$ in Eq. \ref{Eq:lambda defn}. We take the reference $\lambda_{\rm{ref}}=8$ and the coarse $\la_c =\lbrace0.5,1,2,4\rbrace$ and plotted a similar log-log plot of $E_{L_2}^t$ vs. $\Delta t $ in Fig. \ref{fig:Probability_conservation test}(c). The slope of the plot is found to be $\sim -2.12$, inferring a second-order convergence in time step, which is justified as follows. 
As we are using the trapezoidal integration rule where the local truncation error per step is $\sim \mathcal O(\Delta t^3)$. However, the global error is 
\begin{align}\label{Eq:global time error}
    {\rm{global \,\,error}} & \sim N \times   {\rm{local \,\,error}} \nonumber\\
    & \sim \frac{1}{\Delta t} \times \mathcal{O}(\Delta t^3) \sim \mathcal{O}(\Delta t^2). 
\end{align}
This is exactly what we observe in our plot.

\bibliography{scattering}

\end{document}